# In-medium nucleon dynamics at short range: ground-state energy, radii, momentum distribution and spectral functions of few-nucleon systems and complex nuclei⋆

C. Ciofi degli Atti[1], H. Morita[2]

[1]Istituto Nazionale di Fisica Nucleare, Sezione di Perugia, c/o Department of Physics and Geology, University of Perugia, Via A. Pascoli, I–06123, Perugia, Italy
[2]Sapporo Gakuin University, Bunkyo-dai 11, Ebetsu 069–8555, Hokkaido, Japan

**Abstract.** Recent many-body approaches and results concerning the properties of few-nucleon systems and complex nuclei, with particular emphasis to high momentum components and short-range-correlations (SRC), are reviewed and a new approach to the one-nucleon spectral function is presented. The latter is based upon the general and universal properties of two-nucleon momentum distributions as they result from density matrices obtained with realistic many-body ground-state wave functions. The universal features of the momentum distributions are analyzed and shown to follow from a basic property of the nuclear wave function, namely its factorized form at short inter-nucleon distances. The SRC part of the many-body spectral function is expressed in terms of a convolution integral involving ab-initio relative and center-of mass momentum distributions. The obtained convolution spectral function of the three-nucleon systems, featuring both two-and three-nucleon short-range correlations, perfectly agrees in a wide range of momentum and removal energy with the *ab-initio* spectral function, whereas in the case of complex nuclei the integral of the spectral functions (*the momentum sum rule*) reproduces with high accuracy the high momentum part of the one-nucleon momentum distribution, obtained independently from the Fourier transform of the non-diagonal one-body density matrix. Thus, the obtained convolution spectral function for a given nucleus A appears to be a realistic microscopic, parameter-free quantity governed by the features of the underlying two-nucleon interactions.

## 1 Introduction

In this talk recent calculations of the momentum distributions and spectral functions will be presented and discussed, with particular emphasis to the investigation of short range correlations (SRC) and the universal character of the high-momentum part of the nuclear wave function [1,2].

---

⋆ Talk delivered by Claudio Ciofi degli Atti

## 2   The standard many body approach to nuclei: ab-initio and variational calculations

In our analysis we follow the standard description of nuclei as systems composed of point-like non relativistic nucleons interacting via realistic two-nucleon (2N) interactions which explain the wealth of scattering phase shifts, as well as via phenomenological three-nucleon interaction which are requested in order to reproduce the experimental value of the bound three-nucleon system. Thus the many-body problem reduces to the solution of the following non-trivial many-body equation

$$\hat{H}\Psi_n = \left[ -\frac{\hbar^2}{2\,m_N} \sum_i \hat{\nabla}_i^2 + \sum_{i<j} \hat{v}_2(i,j) + \sum_{i<j<k} \hat{v}_3(i,j,k) \right] \Psi_n = E_n\,\Psi_n, \quad (1)$$

where $\Psi_n \equiv \Psi_n(1\ldots A)$, $i \equiv x_i \equiv \{\sigma_i, \tau_i, r_i\}$, $\sum_{i=1}^A r_i = 0$. The most advanced form of realistic 2N interaction is the one of the Argonne family [3]

$$\hat{v}_2(i,j) = \sum_{n=1}^{18} v^{(n)}(r_{ij})\,\hat{\mathcal{O}}_{ij}^{(n)}, \quad (2)$$

with

$$\hat{\mathcal{O}}_{ij}^{(n)} = \left[ 1,\, \sigma_i \cdot \sigma_j,\, \hat{S}_{ij},\, (\mathbf{S} \cdot \mathbf{L})_{ij},\, L^2,\, L^2 \sigma_i \cdot \sigma_j,\, (\mathbf{S} \cdot \mathbf{L})_{ij}^2,\, .. \right] \otimes [1,\, \tau_i \cdot \tau_j] \quad (3)$$

where $\sigma_i$ and $\tau_i$ are Pauli matrices acting in spin and isospin space, respectively, and the most relevant components of (2) are the first six ones. Eq. (1) can be solved exactly (*ab-initio*) only in the case of the $A \leq 4$, whereas for complex nuclei only variational solutions are available, obtained by finding the ground-state wave function $\Psi_0$ which minimizes the expectation value of the Hamiltonian, i.e.

$$\langle \hat{H} \rangle = \frac{\langle \psi_o | \hat{H} | \rangle \psi_o}{\langle \psi_o | \psi_o \rangle} \geq E_o. \quad (4)$$

To-date, three main approaches are pursued to solve the variational problem, namely

1. the variational Monte Carlo (VMC) approach [4], based upon a variational wave function of the type

$$|\Psi_V\rangle = \left(1 + \sum_{i<j<k} U_{ijk}\right) \left[ \hat{\mathcal{S}} \prod_{i<j} \left(1 + U_{ij}^{2-6}\right) \right] \left[ 1 + \sum_{i<j} U_{ij}^{7-8} \right] \left[ \prod_{i<j} f_c(r_{ij}) \right] |\Phi_{MF}\rangle \quad (5)$$

where $U_{ijk}$ and $U_{ij}^n$ are, respectively, 3N and 2N correlation functions (the latter has a spin- isospin dependence corresponding to the different components of the 2N interaction (Eq. (2) )), $f_c(r_{ij})$ is a central correlation function, $|\Phi_{MF}>$ is a mean-field wave function and, eventually, $\hat{\mathcal{S}}$ is a symmetrization operator; with the above general choice of variational function $<\hat{H}>$ is calculated exactly by Monte Carlo numerical integration; the VMC could be applied up to now for nuclei with $A \leq 12$, for which one- and two-nucleon momentum distributions have also been obtained;

2. the cluster variational Monte Carlo (CVCM) [5] which also uses the same wave function as the VMC one, but whereas the integrals involving central correlations are calculated exactly by numerical integrations, the integrals involving non central correlations are evaluated by proper cluster expansions; recently the results for the ground state properties of A = 16 and A = 40, including the one-nucleon momentum distribution, have been released [6];
3. the Fermi Hypernetted Chain (FHNC) method [7,8] and various kinds of linked cluster expansion methods [9–12]; in particular, we have used the so called normalization conserving linked cluster expansion (NCLCE), derived in Ref. [10] and applied for the first time in Ref. [12] with a variational wave function of the type

$$|\Psi_V\rangle = \left[\hat{\mathcal{S}}\prod_{i<j}\left(1+u_{ij}^{2-6}\right)\right]\left[\prod_{i<j}f_c(r_{ij})\right]|\Phi_{MF}\rangle \quad (6)$$

and realistic NN interactions of the Argonne family; it can be seen that the variational wave function of the NCLCE is not as rich as the one used in VMC and CVCM, but we will show that it nonetheless produces, with much less computational efforts, very reasonable and satisfactory results; within the NCLCE the variational problem has been solved and the one- and two-nucleon momentum distributions and spectral functions for iso-scalar nuclei with A ≤ 40 have been obtained [13–18].

In what follows we will briefly discuss the essential points concerning the NCLCE used in our approach. To this aim, consider a generic central operator $\hat{\mathcal{A}}$, so that

$$<\hat{\mathcal{A}}> = \frac{\langle\Psi|\hat{\mathcal{A}}|\Psi\rangle}{\langle\Psi|\Psi\rangle} = \frac{\langle\Phi^{MF}|\prod f_c(r_{ij})\hat{\mathcal{A}}\prod f_c(r_{ij})|\Phi^{MF}\rangle}{\langle\Phi^{MF}|\prod f_c(r_{ij})^2|\Phi^{MF}\rangle}. \quad (7)$$

In this equation the numerator and the denominator contains both linked and unlinked contributions. The latter make the expectation value to diverge with increasing number of particles, a fact which is known even from the theory of quantum fluids [11]. However, by writing

$$f_c(r_{ij})^2 = 1 + \eta(r_{ij}) \quad (8)$$

and expanding the denominator by using $[1+\eta]^{-1} = 1-\eta+\eta^2-\cdots$, the unlinked terms in the numerator exactly cancel out the ones arising from the denominator and a convergent series expansion, containing only linked terms, is obtained in the form

$$\langle\hat{\mathcal{A}}\rangle = \langle\hat{\mathcal{A}}\rangle_o^{MF} + \langle\hat{\mathcal{A}}\rangle_1 + \langle\hat{\mathcal{A}}\rangle_2 + \cdots + \langle\hat{\mathcal{A}}\rangle_n + \cdots, \quad (9)$$

where the subscripts denote the number of $\eta(r_{ij})$ appearing in the given term, with the first term of the expansion representing the MF uncorrelated contribution; the main feature of the expansion can be readily explained in the case of the correlated nuclear density: when the latter is integrated to obtain the number of nucleons, all correlation terms exactly cancel out and the normalization is simply given by the MF term, i.e. the expansion is normalization conserving. To sum up, one of the main aspects of any convergent cluster expansion is to get rid of the explicit presence of the denominator.

All of the three described approaches are genuine realistic many-body approaches, in that the wave function results from the minimization of the expectation value of the Hamiltonian containing realistic NN interactions.

## 3  Comparison of CVMC and NCLCE in coordinate space

In this Section the results of the three methods are briefly compared.

### 3.1  The binding energy and rms radius of $^{16}$O

**Table 1.** The comparison of the binding energy and rms radii of $^{16}$O obtained by the NCLCE [12] and CVMC [6]

| Approach | Mean Field | Potential | (E/A) | (E/A)$_{exp}$ | $\langle r^2 \rangle^{1/2}$ | $(\langle r^2 \rangle^{1/2})_{exp}$ |
|---|---|---|---|---|---|---|
| NCLCE | WS | AV8' | −4.4 | −7.98 | 2.64 | 2.69 |
| CVMC | WS | AV18 | −5.5 | | 2.54 | |
| CVMC | WS | AV18+UIX | −5.15 | | 2.74 | |

The values of the calculated binding energy and rms radius of $^{16}$O are listed in Table 1, which shows that similar results between NCLCE and CVMC are obtained. The CVMC results suggest some repulsive effects by the 3N interaction. Even if the latter is not taken into account in our calculations, it will not quantitatively affect the high momentum content of the nuclear wave function as we shall see in the following.

### 3.2  The operator two-body densities

The two-body densities associated to the first six components of the NN interaction (Eq. (3)) are shown in Fig. 1 which clearly shows the similarity of the results by the two approaches.

## 4  Universality of SRC in coordinate space: the correlation hole and the spin-isospin content of the ground-state wave function

### 4.1  The two-body densities $\rho_{(2)}$ in few-nucleon systems and complex nuclei

The correlation hole is the strong reduction at $r = |\mathbf{r}_1 - \mathbf{r}_2| < 1.0$ fm of the realistic 2N density distribution with respect to the mean-field predictions. It can be seen from Fig. 2 that, apart from a normalization factor, the correlation hole is universal (A-independent) and similar to the hole in the deuteron density.

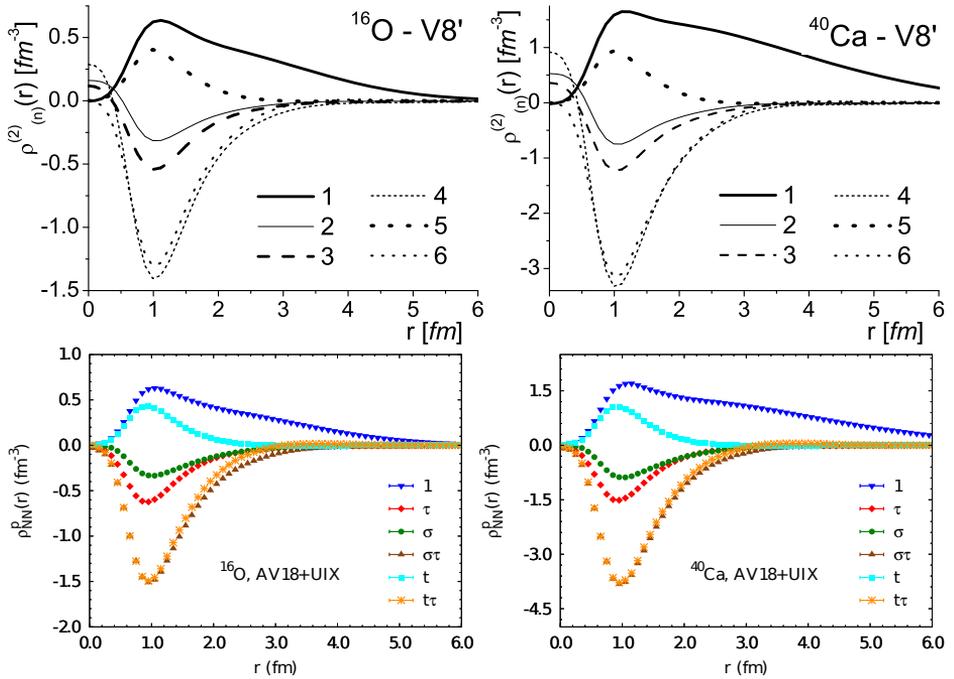

**Fig. 1.** The two-body densities corresponding to the first six components of the AV18 interaction. Top: Ref. [12]; Bottom: Ref. [6].

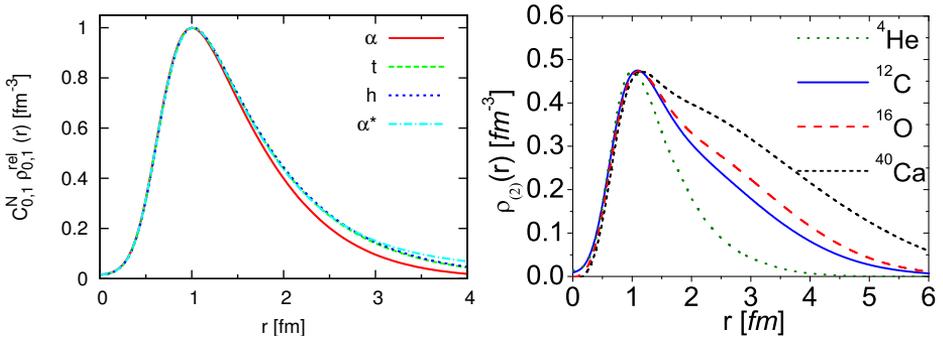

**Fig. 2.** The calculated two-body densities in various nuclei showing the universal correlation hole in the region $r = |\mathbf{r}_1 - \mathbf{r}_2| \leq 1$ fm. Left: after Ref. [20]; Right: ref. [19]

### 4.2  On the effects of 3N interaction on the two-body operator densities

As illustrated in Fig. 3, 3N interaction does not practically affect the correlation hole.

### 4.3  The spin-isospin content of the nuclear wave function

Two nucleon systems in nuclei have to satisfy Pauli principle: $L + S + T -$ odd. In Table 2 the number of NN pairs in various spin-isospin (ST) are listed and a very good agreement between various many-body approaches can be noticed. Moreover a large number of odd orbital momentum states are predicted to be present in the ground state wave function, particularly in heavy nuclei.

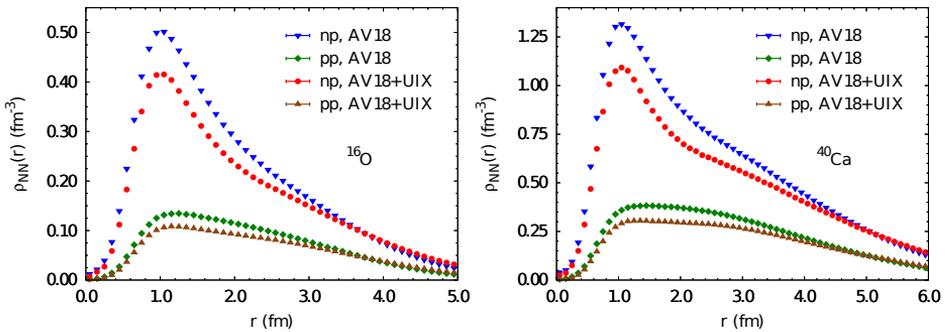

**Fig. 3.** The two-body densities of $^{16}$O and $^{40}$Ca from Ref. [6] including (AV18+UIX) and omitting (AV18) the three-nucleon potential UIX. (After Ref. [6])

**Table 2.** The number of pairs $N^A_{(ST)}$ in various spin-isospin states in the independent particle model (IPM) and taking into account SRC within many-body theories with realistic interactions.(After Ref. [16])

| Nucleus | | (ST) (10) | (01) | (00) | (11) |
|---|---|---|---|---|---|
| $^2$H | | 1 | - | - | - |
| $^3$He | IPM | 1.50 | 1.50 | - | - |
| | SRC [16] | 1.488 | 1.360 | 0.013 | 0.139 |
| | SRC [21] | 1.50 | 1.350 | 0.01 | 0.14 |
| | SRC [20] | 1.489 | 1.361 | 0.011 | 0.139 |
| $^4$He | IPM | 3 | 3 | - | - |
| | SRC [16] | 2.99 | 2.57 | 0.01 | 0.43 |
| | SRC [21] | 3.02 | 2.5 | 0.01 | 0.47 |
| | SRC [20] | 2.992 | 2.572 | 0.08 | 0.428 |
| $^{16}$O | IPM | 30 | 30 | 6 | 54 |
| | SRC [16] | 29.8 | 27.5 | 6.075 | 56.7 |
| | SRC [21] | 30.05 | 28.4 | 6.05 | 55.5 |
| $^{40}$Ca | IPM | 165 | 165 | 45 | 405 |
| | SRC [16] | 165.18 | 159.39 | 45.10 | 410.34 |

## 5 SRC in momentum space

### 5.1 One-nucleon momentum distribution (1NMD)

The 1NMD is the Fourier transform of the 1N non diagonal density matrix

$$\rho(\mathbf{r}_1, \mathbf{r}'_1) = \int \Psi^*_0(\mathbf{r}_1, \mathbf{r}_2 \ldots, \mathbf{r}_A) \Psi_0(\mathbf{r}'_1, \mathbf{r}_2 \ldots, \mathbf{r}_A) \prod_{i=2}^A d\mathbf{r}_i \quad (10)$$

namely

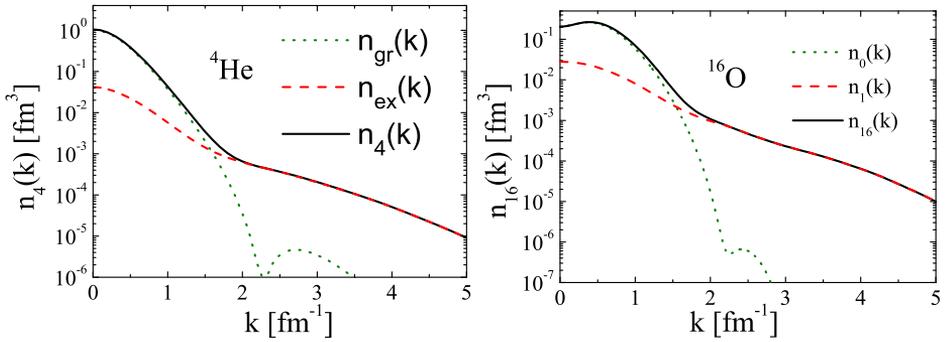

**Fig. 4.** The mean-field ($n_{gr}$ and $n_0$) and SRC ($n_{ex}$ and $n_1$) contribution in the one-nucleon momentum distribution of $^4$He and $^{16}$O (After Ref. [16])

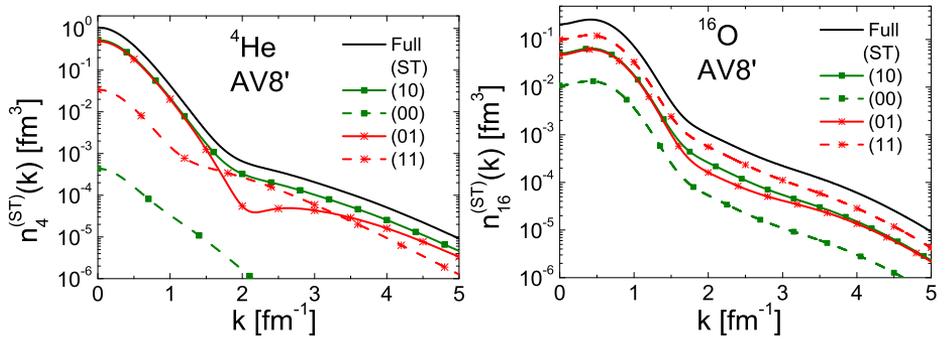

**Fig. 5.** The ST contributions to the one-nucleon momentum distribution of $^4$He and $^{16}$O (After Ref. [16])

$$n(\mathbf{k}_1) = \int e^{-i\,\mathbf{k}_1\cdot(\mathbf{r}_1-\mathbf{r}'_1)}\rho(\mathbf{r}_1,\mathbf{r}'_1)d\mathbf{r}_1\,d\mathbf{r}'_1, \qquad (11)$$

which can be written as

$$n(\mathbf{k}_1) = n_{MF}(\mathbf{k}_1) + n_{SRC}(\mathbf{k}_1). \qquad (12)$$

Where the first term is the MF distribution renormalized by the spectroscopic factors [22,8] and the second part describes SRC. The results for $^4$He and $^{16}$O are shown in Fig. 4 whereas Fig. 5 shows the the ST contributions to the one-nucleon momentum distribution.

### 5.2  Effects of 3N interaction on the 1NMD

It turns out, as shown in Fig. 6, that 3N interaction only slightly affects the high momentum content of the ground state momentum distributions.

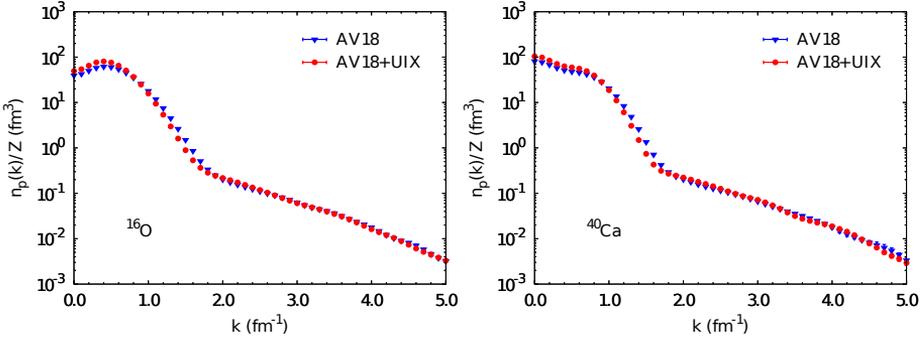

**Fig. 6.** The effects of three-nucleon forces on the one-nucleon momentum distributions (After Ref. [6])

### 5.3 The two-nucleon momentum distributions (2NMD)

By introducing the relative (*rel*) and center-of-mass (*c.m.*) momenta

$$\mathbf{k}_{rel} = \frac{1}{2}(\mathbf{k}_1 - \mathbf{k}_2) \qquad \mathbf{K}_{c.m.} = \mathbf{k}_1 + \mathbf{k}_2 \qquad (13)$$

the following two-nucleon momentum distributions can be introduced:

1. The total two-nucleon momentum distribution

$$n(\mathbf{k}_1, \mathbf{k}_2) = n(\mathbf{k}_{rel}, \mathbf{K}_{c.m.}) = n(k_{rel}, K_{c.m.}, \theta)$$
$$= \frac{1}{(2\pi)^6} \int d\mathbf{r} d\mathbf{r}' d\mathbf{R} d\mathbf{R}' e^{-i\mathbf{K}_{c.m.} \cdot (\mathbf{R}-\mathbf{R}')} e^{-i\mathbf{k}_{rel}(\mathbf{r}-\mathbf{r}')} \rho^{(2)}(\mathbf{r},\mathbf{r}';\mathbf{R},\mathbf{R}') \quad (14)$$

2. The back-to-back two-nucleon momentum distribution

$$n(\mathbf{k}_{rel}, \mathbf{K}_{c.m.} = 0) \qquad (15)$$

3. The relative two-nucleon momentum distribution

$$n_{rel}(\mathbf{k}_{rel}) = \int n(\mathbf{k}_{rel}, \mathbf{K}_{c.m.}) d\mathbf{K}_{c.m.} \qquad (16)$$

4. The c.m. momentum distribution

$$n_{c.m.}(\mathbf{K}_{c.m.}) = \int n(\mathbf{k}_{rel}, \mathbf{K}_{c.m.}) d\mathbf{k}_{rel} \qquad (17)$$

If $n(k_{rel}, K_{c.m.}, \theta)$ is $\theta$ independent, it means that $n(\mathbf{k}_{rel}, \mathbf{K}_{c.m.}) = n(\mathbf{k}_{rel}) n(\mathbf{K}_{c.m.})$ i.e. the relative and c.m. motions factorize. Factorization has fundamental consequences on the structure of the high momentum components in nuclei.

### 5.4 Comparison of the 1NMD calculated within CVMC and NCLCE

In Fig. 7 the 1NMD of $^{16}$O and $^{40}$Ca calculated within the NCLCE and CVMC approaches are compared and a satisfactory agreement can be seen.

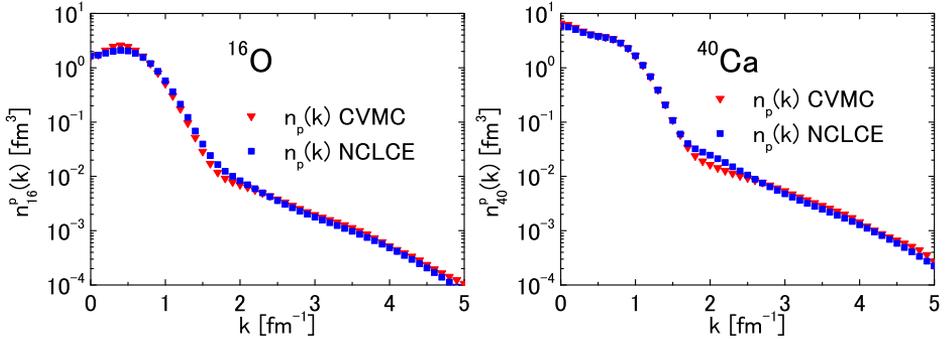

**Fig. 7.** The one-nucleon momentum distributions of $^{16}$O and $^{40}$Ca from Ref. [6] compared with the momentum distribution obtained in Ref. [12,16].

### 5.5 The proton-neutron and proton-proton momentum distributions

In Fig. 8 (left) we show the pn and pp 2NMD (Eq. (14)) for $^4$He, calculated *vs* $k_{rel}$, in correspondence of several values of the c.m. momentum and two different values of the angle θ. It can be seen that at high values of $k_{rel}$ independence upon the angle really occurs, which is evidence of factorization of the 2NMD. The same trend is observed for complex nuclei. It can also be seen that, thanks to the effect of the tensor force, the pn distribution are much larger than the pp ones; the difference however is smaller in the integrated momentum distributions shown in Fig. 8 (right), where a satisfactory agreement between NCLCE and VMC can be seen.

## 6 Universality and factorization of the 2NMD

At large $k_{rel}$ ($\geq 2\,\mathrm{fm}^{-1}$) and small $K_{c.m.}$ ($\leq 1\,\mathrm{fm}^{-1}$) the 2NMD typically factorizes; in particular in the case of pn pairs one has

$$n_A^{pn}(\mathbf{k}_{rel}, \mathbf{K}_{c.m.}) \Longrightarrow n_A^{pn}(\mathbf{k}_{rel}, \mathbf{K}_{c.m.}) \simeq C_A^{pn} n_D(k_{rel}) n_{c.m.}^{pn}(K_{c.m.}) \quad (18)$$

where $n_D$ is the deuteron momentum distribution. By plotting the quantity

$$\frac{n_A^{pn}(k_{rel}, K_{c.m.} = 0)}{n_{c.m.}^{pn}(K_{c.m.} = 0) n_D(k_{rel})}, \quad (19)$$

the value of $C_A$ can be obtained when the ratio becomes a constant. This is shown in Fig. 9 and the obtained values of $C_A$ are listed in Table 3.

Moreover, if factorization holds one should have at $k_{rel} > k_{rel}^- \simeq 2\,\mathrm{fm}^{-1}$

$$R_{fact/exact}^{pn} \equiv \frac{C_A^{pn} n_D(k_{rel}) n_{c.m.}(K_{c.m.})}{n_A^{pn}(k_{rel}, K_{cm}, \theta)} \Rightarrow 1 \quad (20)$$

which is indeed confirmed in Fig. 10, which represents another evidence of the universality of SRC.

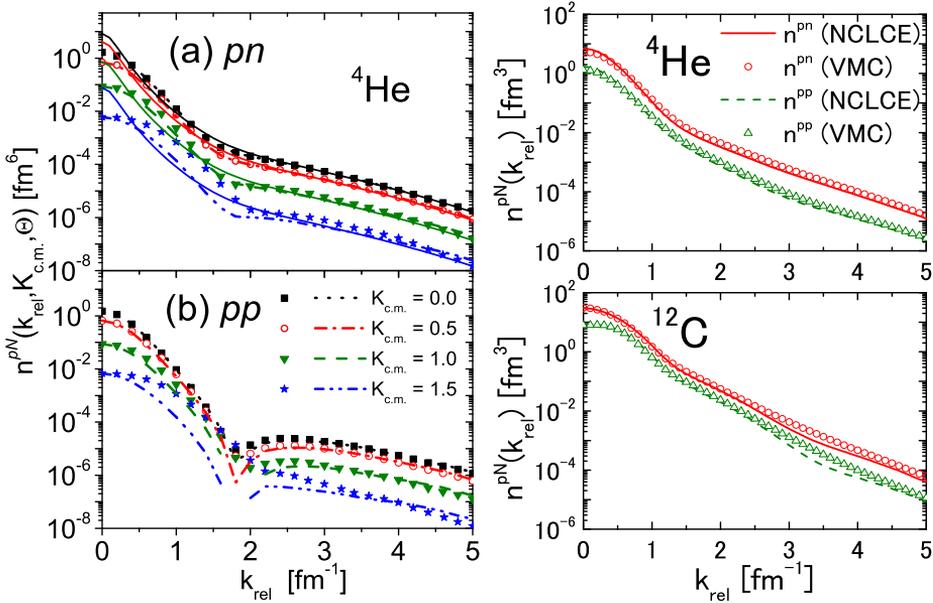

**Fig. 8.** Left panel: the pn and pp two-body momentum distribution for $A = 4$ (after Ref. [14]). Right panel: the pn and pp relative momentum distributions (Eq. (16)) in $^4$He and $^{12}$C calculated by the NCLCE (lines) [13] and the VMC (symbols) [4] (after Ref. [18]).

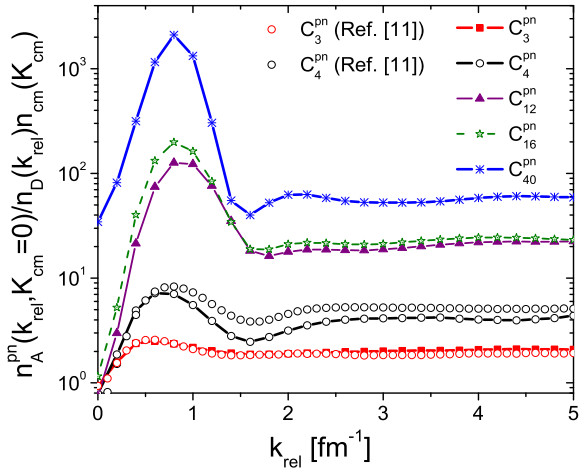

**Fig. 9.** The quantity given in Eq. (19). (After Ref. [13].)

**Table 3.** The values of the constant $C_A^{pn}$ appearing in Eq. (18) extracted from Fig. 9, with error determined according to the following expression: $C_A^{pn} = \left((C_A^{pn})^{Max} + (C_A^{pn})^{Min}\right)/2 \pm \left((C_A^{pn})^{Max} - (C_A^{pn})^{Min}\right)/2$, where $(C_A^{pn})^{Max}$ and $(C_A^{pn})^{Min}$ are determined in the region of $k_{rel} \geq 3.0 \text{fm}^{-1}$. The values in brackets have been obtained using the VMC wave function of Ref. [4]. (After Ref. [13])

| $^2$H | $^3$He | $^4$He | $^6$Li | $^8$Be | $^{12}$C | $^{16}$O | $^{40}$Ca |
|---|---|---|---|---|---|---|---|
| 1.0 | $2.0 \pm 0.1$ | $4.0 \pm 0.1$ | – | – | $20 \pm 1.6$ | $24 \pm 1.8$ | $60 \pm 4.0$ |
| 1.0 | $(2.0 \pm 0.1)$ | $(5.0 \pm 0.1)$ | $(11.1 \pm 1.3)$ | $(16.5 \pm 1.5)$ | (–) | (–) | (–) |

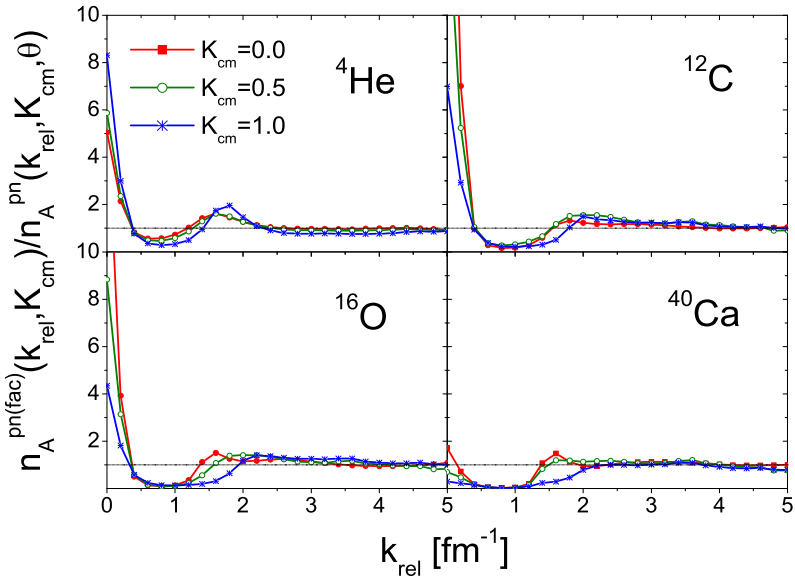

**Fig. 10.** The quantity $R_{fact/exact}^{pn} \equiv C_A^{pn} n_D(k_{rel}) n_{c.m.}(K_{c.m.}) / n_A^{pn}(k_{rel}, K_{cm}, \theta)$. (After Ref. [13].)

So far about pn pairs, but what about pp pairs? It turns out also the pp 2NMD factorizes, exhibiting, moreover, a high degree of universality. This is illustrated in Fig. 11 where it can be seen that the pp distributions for a nucleus $A$ scale indeed to the pp distributions in $^4$He.

To sum up, we would like to stress that the 2NMD exhibit, in a very clear way, their universality character. It should be pointed out that the region of factorization depends upon the value of $K_{c.m.}$. Indeed the factorization region starts at values of the relative momentum $k_{rel}$ given by

$$k_{rel} > k_{rel}^-(K_{c.m.}) \simeq C_1 + C_2 \Phi_A(K_{c.m.}), \qquad (21)$$
$$C_1 \simeq 1 \, \text{fm}^{-1} \quad C_2 \simeq 0.5 \, \text{fm}^{-1} \quad \Phi_A = |\mathbf{K}_{c.m.}|,$$

which is derived from the analysis of Fig 12 in the case of $A = 4$ and this is also applicable for complex nuclei [18]. This dependence of the factorization region upon the value of $K_{c.m.}$ introduces a constraint in the derivation of the spectral function, to be discussed in the next Section.

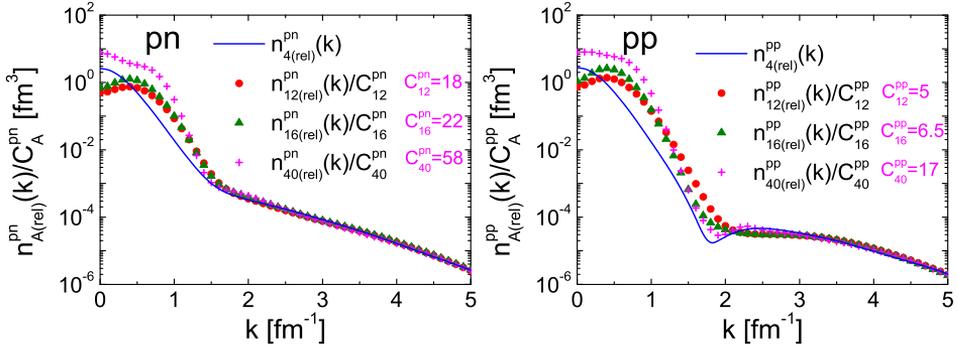

**Fig. 11.** The quantity $n_A^{pN}(k_{rel}, K_{c.m.} = 0)/(C_A^{pN} n_{c.m.}(K_{c.m.} = 0)) \simeq n_4^{pN}(k_{rel})$ when $k_{rel} > 1.5(pn)$ fm$^{-1}$ and $2.5(pp)$ fm$^{-1}$.

## 7 Factorization and the convolution model of the one-nucleon spectral function of few-nucleon systems and complex nuclei

The nucleon spectral function (SF) is one of the basic nuclear structure quantity in the description $e - A$ and $\nu - A$ scattering processes. It is defined as follows

$$P_A^N(\mathbf{k}_1, E) = \frac{1}{2J+1} \sum_{M,\sigma_1} \langle \Psi_A^{JM} | a_{\mathbf{k}_1\sigma_1}^\dagger \delta(E - (\hat{H}_A - E_A)) a_{\mathbf{k}_1\sigma_1} | \Psi_A^{JM} \rangle \quad (22)$$

$$= \frac{1}{(2J+1)} (2\pi)^{-3} \sum_{M,\sigma_1} \sum_f \left| \int d\mathbf{r}_1 e^{i\mathbf{k}_1 \cdot \mathbf{r}_1} G_f^{M\sigma_1}(\mathbf{r}_1) \right|^2 \delta(E - (E_{A-1}^f - E_A)),$$

$$G_f^{M\sigma_1}(\mathbf{r}_1) = \langle \chi_{\sigma_1}^{1/2}, \Psi_{A-1}^f(\{x\}_{A-1}) | \Psi_A^{JM}(\mathbf{r}_1, \{x\}_{A-1}) \rangle. \quad (23)$$

The integral of the SF over the removal energy E (the momentum sum rule) provides the nucleon momentum distribution

$$n_A^N(k_1) = \int P_A^N(k_1, E) dE. \quad (24)$$

The SF can be represented in the following form

$$P_A^N(k_1, E) = P_{MF}^N(k_1, E) + P_{SRC}^N(k_1, E), \quad (25)$$

where $P_{MF}^N$ represents the mean-field SF with occupied shell-model states $\alpha$, below the Fermi level, renormalized by the spectroscopic factors $Z_\alpha < 1$ [8,22], and $P_{SRC}^N$ is the correlated SF populating states above the Fermi levels [1]. In particular one has

$$P_{MF}^N(k_1, E) = \frac{1}{(2\pi)^3(2J+1)} \sum_{M,\sigma,f \leq F} \left| \int e^{i\mathbf{k}_1 \cdot \mathbf{r}_1} G_f^{M\sigma}(\mathbf{r}_1) d\mathbf{r}_1 \right|^2 \delta(E - E^f), \quad (26)$$

$$P_{SRC}^N(k_1, E) = \frac{1}{(2\pi)^3(2J+1)} \sum_{M,\sigma,f > F} \left| \int e^{i\mathbf{k}_1 \cdot \mathbf{r}_1} G_f^{M\sigma}(\mathbf{r}_1) d\mathbf{r}_1 \right|^2 \delta(E - E^f). \quad (27)$$

---

[1] Different notations are used by different authors, e.g.: $P_A^N = P_0 + P_1$, $P_A^N = P_{gr} + P_{ex}$, etc.

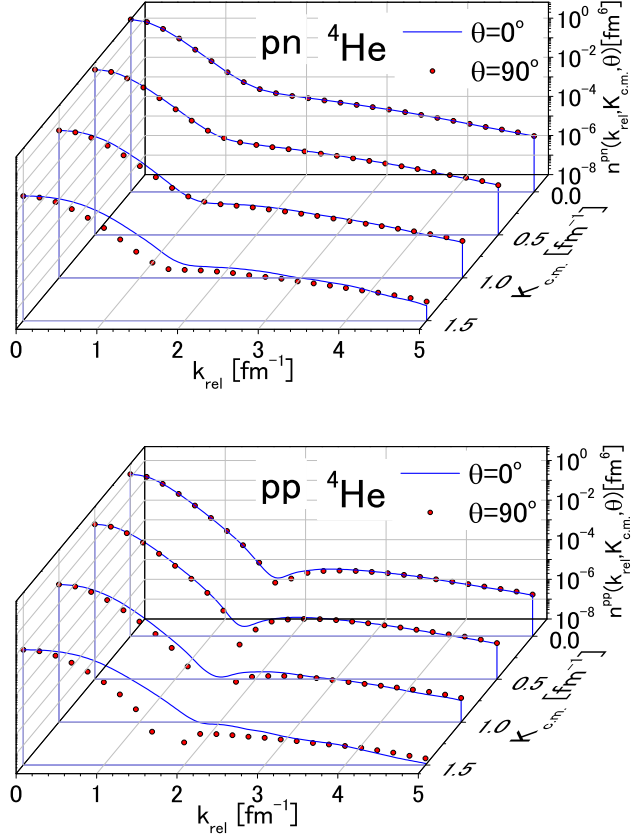

**Fig. 12.** The pn and pp two-nucleon momentum distributions in $^4$He, $n^{pN}(k_{rel}, K_{c.m.}, \theta)$, obtained in Ref. [14] in correspondence of several values of $K_{c.m.}$ and two values of the angle $\theta$ between $\mathbf{K}_{c.m.}$ and $\mathbf{k}_{rel}$ (After Ref. [18]). The region of $k_{rel}$ where the value of $n^{pN}(k_{rel}, K_{c.m.}, \theta)$ is independent of the angle determines the region of factorization of the momentum distributions, i.e. $n^{pN}(k_{rel}, K_{c.m.}, \theta) \to n_{rel}^{pN}(k_{rel}) n_{c.m.}^{pN}(K_{c.m.})$. It can be seen that the region of factorization starts at values of $k_{rel} = \bar{k}_{rel}^-$, which increases with increasing values of $K_{c.m.}$, i.e. $\bar{k}_{rel}^- = \bar{k}_{rel}^-(K_{c.m.})$; because of the dependence of $\bar{k}_{rel}^-$ upon $K_{c.m.}$, a constraint on the region of integration over $K_{c.m.}$ arises from Eq. (21).

The exact relation between the one- and two-nucleon momentum distributions ($N_1 \neq N_2$)

$$n_A^{N_1}(\mathbf{k}_1) = \frac{1}{A-1}\left[\int n_A^{N_1 N_2}(\mathbf{k}_1, \mathbf{k}_2)\, d\mathbf{k}_2 + 2\int n_A^{N_1 N_1}(\mathbf{k}_1, \mathbf{k}_2)\, d\mathbf{k}_2\right] \quad (28)$$

becomes, *in the factorization region*, as follows

$$n_A^{SRC}(\mathbf{k}_1) = \left[\int n_{rel}^{N_1 N_2}(|\mathbf{k}_1 - \frac{\mathbf{K}_{c.m.}}{2}|) n_{c.m.}^{N_1 N_2}(\mathbf{K}_{c.m.})\, d\mathbf{K}_{c.m.}\right.$$
$$\left. + 2\int n_{rel}^{N_1 N_1}(|\mathbf{k}_1 - \frac{\mathbf{K}_{c.m.}}{2}|) n_{c.m.}^{N_1 N_1}(\mathbf{K}_{c.m.})\, d\mathbf{K}_{c.m.}\right] \equiv n_{ex}^{n_1}(\mathbf{k}_1), \quad (29)$$

so that the correlated part of the nucleon spectral function is

$$P_{SRC}^{N_1}(\mathbf{k}_1, E) = \sum_{N_2=p,n} C_{N_1 N_2} \int n_{rel}^{N_1 N_2}(|\mathbf{k}_1 - \frac{\mathbf{K}_{c.m.}}{2}|) n_{c.m.}^{N_1 N_2}(\mathbf{K}_{c.m.}) d\mathbf{K}_{c.m.}$$
$$\times \delta\left(E - E_{thr} - \frac{A-2}{2m_N(A-1)}\left[\mathbf{k}_1 - \frac{(A-1)\mathbf{K}_{c.m.}}{A-2}\right]^2\right), \qquad (30)$$

where $C_{N_1=N_2} = 2$, $C_{N_1 \neq N_2} = 1$. This is the *convolution formula* of the SF. It should be remembered that a similar convolution formula has been previously obtained in Ref. [26], and frequently applied over the years to the description of various kind of processes. However it should also be stressed that the model of Ref. [26]: (i) is based upon the phenomenological relative and c.m. distributions for pn and pp pairs, since in 1996 the realistic pn and pp momentum distributions were unknown; (ii) for the same reason the constraint on the values of $K_{c.m.}$ imposed by the variation of the factorization region with increasing values of $K_{c.m.}$, was not taken into account. For these reasons it appears to be important to calculate the convolution SF using realistic many-body relative and c.m. distributions and taking into account the constraint on $K_{c.m.}$; Using a realistic c.m. momentum distribution (having both low and high momentum components) one can also investigate the effects of both 2N and 3N SRC which has been recently done in the case of $^3$He [17].

### 7.1  The convolution SF of $^3$He

There exist in the literature various *ab initio* calculations of the SF of the three-nucleon system[23]; we will use in what follows the most recent ones obtained with the AV18 interaction [24]. The convolution neutron and proton spectral functions are as follows:

$$P_{SRC}^n(\mathbf{k}_1, E) = \int n_{rel}^{np}(|\mathbf{k}_1 - \frac{\mathbf{K}_{c.m.}}{2}|) \, n_{c.m.}^{np}(\mathbf{K}_{c.m.}) \, d\mathbf{K}_{c.m.} \delta\left(E - E_{thr} - \frac{1}{4m_N}[\mathbf{k}_1 - 2\mathbf{K}_{c.m.}]^2\right),$$

$$P_{SRC}^p(\mathbf{k}_1, E) = P_{SRC}^n(\mathbf{k}_1, E) + 2\int n_{rel}^{pp}(|\mathbf{k}_1 - \frac{\mathbf{K}_{c.m.}}{2}|) n_{c.m.}^{pp}(\mathbf{K}_{c.m.}) \, d\mathbf{K}_{c.m.} \delta\left(E - E_{thr} - \frac{1}{4m_N}[\mathbf{k}_1 - 2\mathbf{K}_{c.m.}]^2\right)$$

In Fig. 13 the (parameter-free) convolution formula is compared with the *ab-initio* one and a very satisfactory agreement is found, except at very high removal energies, a region which is anyway difficult to reach experimentally.

### 7.2  The convolution SF of $^4$He and complex nuclei

In Fig. 14 we show the convolution spectral function of $^4$He and complex nuclei with the separate contributions of pn and pp SRC. Finally in Fig. 15 we compare the microscopic convolution formula with the results of Ref. [26]. It can be seen that the results obtained in 1996 are to a large extent confirmed by the present calculations.

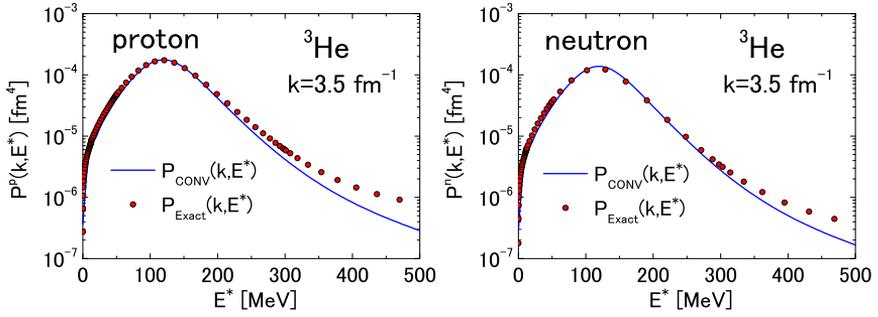

**Fig. 13.** The neutron and proton spectral function of $^3$He calculated within the convolution formula compared with the *ab-initio* spectral function. [17,18]

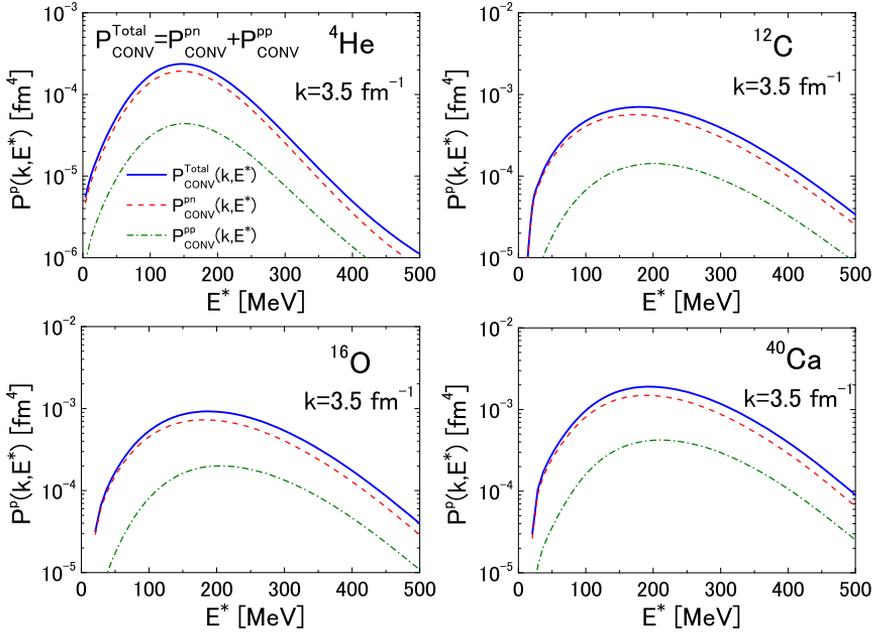

**Fig. 14.** The spectral function of $^4$He, $^{12}$C, $^{16}$O and $^{40}$Ca obtained by the convolution formula showing the contributions from pn and pp SRC (After Ref. [18]).

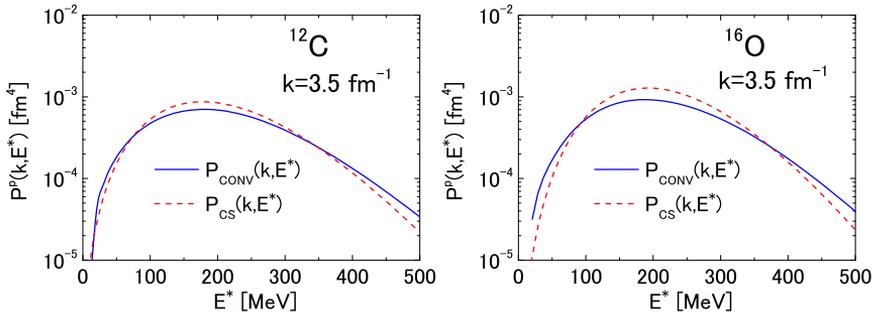

**Fig. 15.** Comparison of the microscopic convolution formula with the effective convolution model of Ref. [26] (CS) (After Ref. [18]).

### 7.3 The momentum sum rule

In Fig. 16 we show the results of the calculation of the momentum sum rule (Eq.(24)) It can be seen that the convolution formula does fully satisfy the momentum sum rule.

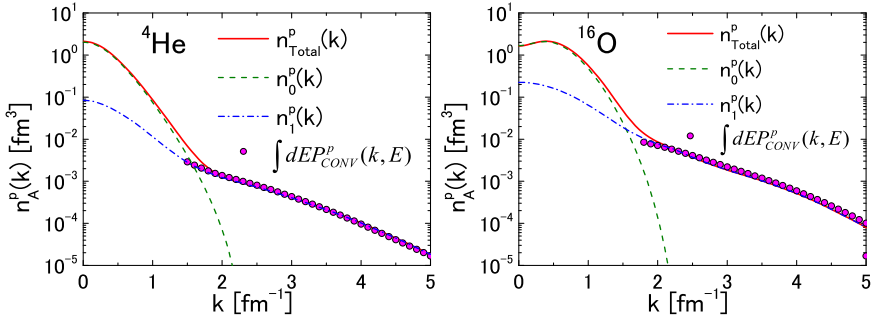

**Fig. 16.** The SRC Momentum sum rule $n_{SRC}(k) \equiv n_1(k) = \int_0^\infty P(k, E^*) dE^*$ in $^4$He and $^{16}$O. The full line represents the total momentum distribution shown in Ref. [16] with the dashed and dot-dashed curves corresponding to the mean-field and SRC contributions, respectively. The full dots represent the the SRC momentum distribution obtained by integrating the the SRC convolution SF. It can be seen that the momentum sum rule is exactly satisfied by the convolution formula. (After Ref. [18])

### 7.4 About the convergence of the momentum sum rule

In Fig.17 we show the quantity

$$n_{E^+}(k) = \int_0^{E^+} P_{SRC}(k, E) \, dE \qquad (31)$$

This figure, in agreement with Fig. 13 of Ref. [26], illustrates that in order to obtain the momentum distributions at high value of the momentum high values of the removal energies have to be considered.

## 8 The origin of factorization of momentum distributions

The factorized structure of two nucleon momentum distributions results from a general property of the nuclear many-body wave function, namely its factorized form at short internucleon distances [26,25], namely

$$\lim_{r_{ij} \to 0} \Psi_0(\{\mathbf{r}\}_A) \simeq \qquad (32)$$
$$\hat{\mathcal{A}} \Big\{ \chi_0(\mathbf{R}_{ij}) \sum_{n, f_{A-2}} a_{0,n,f_{A-2}} \Big[ \Phi_n(\mathbf{x}_{ij}, \mathbf{r}_{ij}) \oplus \Psi_{f_{A-2}}(\{\mathbf{x}\}_{A-2}, \{\mathbf{r}\}_{A-2}) \Big] \Big\}.$$

Factorized wave functions have been introduced in the past as physically sound approximations of the unknown nuclear wave function (see e.g. [27]), without however providing any evidence of the validity of such an approximation due to the lack, at that time, of realistic solutions of the nuclear many-body problem. These however became recently available and the validity of the factorization approximation could be quantitatively checked. Indeed the factorization property of realistic many-body wave functions has been proved to hold in the case of *ab initio* wave functions of few-nucleon systems [28], complex nuclei [13] and nuclear matter [29].

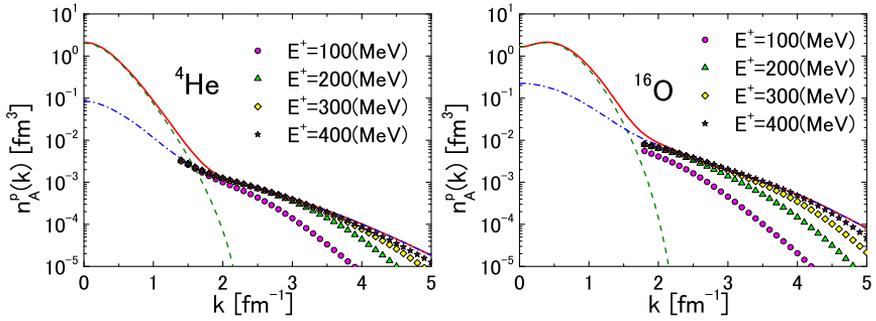

**Fig. 17.** The convergence of the momentum sum rule $n_{SRC}(k) \equiv n_1(k) = \int_0^{E^+} P(k, E^*) \, dE^*$. The partial momentum sum rule corresponding to increasing value of $E^+$. It can be seen that in order to obtain the correct momentum distributions in the region $k \geq 4 \, \text{fm}^{-1}$ it is necessary to integrate the SF up to $E^+ \simeq 400 \, \text{MeV}$. Full, dashed and dot-dashed curves as in Fig. 16. (After Ref. [18])

## 9  Summary and conclusions

- A canonical many-body variational approach has been followed: 1. the 2N interaction (Argonne family) and the general form of the many-body wave function $\Psi_A$ embodying central, spin,isospin, tensor etc have been chosen; 2. the minimization of $< \Psi_A | \hat{H} | \Psi_A > < \Psi_A | \Psi_A >^{-1}$ has been performed and the parameters entering the wave function were determined; 3. the one, $n_A(k_1)$ and two, $n_A^{NN}(k_{rel}, K_{c.m.}, \theta)$, momentum distributions, free of any adjustable parameter, have been calculated.
- The one-nucleon momentum distribution at $k > k_F$ shows high momentum contents which cannot be reconciled with Hartree-Fock or Brueckner-Hartree-Fock type descriptions of nuclei;
- It is demonstrated that, starting from a certain value of the relative momentum (depending upon the value of the c.m. momentum), the two-nucleon momentum distributions factorizes, i. e. obeys the relation $n_A^{N_1 N_2}(k_{rel}, K_{c.m.}, \theta) \simeq n_{rel}^{N_1 N_2}(k_{rel}) n_{c.m.}^{N_1 N_2}(K_{c.m.})$ and the region of factorizations and the explicit form of $n_{rel}^{N_1 N_2}(k_{rel})$ and $n_{c.m.}^{N_1 N_2}(K_{c.m.})$ have been obtained;
- The two-nucleon momentum distributions in the SRC region exhibits universality, i.e., apart from a scaling factor, they are A independent; in case of $pn$ pairs their $k_{rel}$ behavior is governed by the deuteron momentum distribution and their amplitude by the c.m. momentum distribution of the pair; as for the pp distributions their $k_{rel}$ dependence in a complex nucleus is governed by the pp momentum distribution in $^4$He and their amplitude, as in the case of pn pairs, by the c.m. distribution.
- Using the above properties a model-independent parameter-free, fully microscopic spectral function has been obtained.